\documentclass[aps,prc,floatfix]{revtex4}
\usepackage{amsmath}
\usepackage{bm}
\usepackage{psfig}
\usepackage{graphicx}
\usepackage{dcolumn}

\begin{document}

\newcommand{\etal}{{\it et al.\/ }}
\newcommand{\GeVsq}{(GeV/$c$)$^2$ }

\title{Influence of the Dirac sea on proton electromagnetic knockout}
\author{James J. Kelly}
\affiliation{Department of Physics, University of Maryland, 
College Park, MD 20742}
\date{January 28, 2005}

\begin{abstract}
We use the relativistic distorted-wave impulse approximation (RDWIA)
to study the effects of negative-energy components of Dirac wave
functions on the left-right asymmetry for $(e,e^\prime p)$ 
reactions on $^{16}$O with $0.2 \leq Q^2 \leq 0.8$ and $^{12}$C
with $0.6 \leq Q^2 \leq 1.8$ (GeV/$c$)$^2$.
Spinor distortion is more important for the bound state than
for the ejectile and the net effect decreases with $Q^2$.
Spinor distortion breaks Gordon equivalence and the data favor
the CC2 operator with intermediate coupling to the sea.
The left-right asymmetry for $Q^2 \lesssim 1.2$ (GeV/$c$)$^2$
is described well by RDWIA calculations, but at 
$Q^2 = 1.8$ (GeV/$c$)$^2$ the observed variation with missing
momentum is flatter than predicted.
\end{abstract}

\maketitle

\section{Introduction}

One of the interesting features of nucleon electromagnetic knockout reactions
is the sensitivity of some observables to negative-energy states in the 
Dirac sea.
When a Dirac equation with scalar and vector potentials, $S$ and $V$, is 
transformed to an equivalent Schr\"odinger equation, the effective spin-orbit 
potential is proportional to $S-V$.
Within a nucleus, $S-V$ reduces the effective mass, thereby reducing the
upper and enhancing the lower components of the Dirac wave function.
Modification of the relationship between lower and upper components 
requires admixture of negative-energy Dirac states.
Observables sensitive to coupling between lower and upper components can
then reveal the role of negative-energy states.

Several recent papers have investigated the effects of spinor distortion
in some detail 
\cite{Caballero98a,Caballero98b,Kelly99b,Udias99,Udias01,Vignote04,Martinez04}.
In this short paper we will explore the $Q^2$ dependence of such effects
using data for $^{16}$O and $^{12}$C$(e,e^\prime p)$ in quasiperpendicular
kinematics for $Q^2$ between about 0.2 and 1.8 (GeV/$c$)$^2$.
The data for $^{16}$O have already been analyzed in considerable detail 
and the enhancement of the left-right asymmetry for 
$Q^2 = 0.8$ (GeV/$c$)$^2$  has been interpreted as 
evidence for spinor distortion \cite{Gao00,Fissum04}.
Here we use the availability of both $1p_{3/2}$ and $1p_{1/2}$ magnetic
substates to illustrate the sensitivity to several important aspects
of spinor distortion. 
The recent $^{12}$C$(e,e^\prime p)$ data from Dutta \cite{Dutta03} 
reaches larger $Q^2$ but were acquired to study nuclear transparency and 
were analyzed using either nonrelativistic or factorized calculations that 
omit spinor distortion.
Here we use relativistic distorted-wave calculations to show that
the left-right asymmetry in the reduced cross section for $Q^2 = 0.6$, 1.2, 
and 1.8 (GeV/$c$)$^2$ is sensitive to spinor distortion also.
We consider both semi-inclusive data and bins centered upon the
$1p_{3/2}$ and $1s_{1/2}$ shells.

The model is described briefly in Sec. \ref{sec:model}.
Data for $^{16}$O and $^{12}$C are examined in Secs. \ref{sec:16O} and
\ref{sec:12C}, respectively.
Our conclusions are summarized in Sec. \ref{sec:conclusions}.

\section{Model}
\label{sec:model}

In this section we outline a series of approximations to the relativistic 
distorted-wave impulse approximation (RDWIA) that facilitate exploration 
of the effects of the Dirac sea and comparisons with nonrelativistic
approaches (NRDWIA).
Further details may be found in Ref. \cite{Fissum04}.
All calculations treat electron distortion in the $q_\text{eff}$ 
approximation and use MMD form factors \cite{MMD} and Coulomb gauge.
Most calculations use the CC2 current operator unless noted
otherwise.
We consider Dirac-Hartree wave functions from the original Horowitz and
Serot (HS) analysis \cite{Horowitz86} and NLSH wave functions from
Sharma \etal \cite{Sharma93}.
Optical potentials for Dirac phenomenology were taken from the 
global analysis by Cooper \etal \cite{Cooper93}.

In RDWIA a nuclear matrix element of the single-nucleon electromagnetic 
current for the $A(e,e^\prime N)B$ reaction takes the form
\begin{equation}
{\cal J}^\mu =  
\int d^3 r \; \exp{(i \bm{t}\cdot \bm{r})}
\langle \Psi^{(-)}(\bm{p}^\prime,\bm{r}) | 
\gamma^0 \Gamma^\mu(\bm{p}^\prime,\bm{p}^\prime-\bm{q}) |
\Phi(\bm{r}) \rangle
\label{eq:currme} 
\end{equation}
where $\Gamma^\mu$ is the vertex function,
$\bm{q}$ is the momentum transfer, 
$\bm{p}^\prime$ is the ejectile momentum, and 
$\bm{t} = E_B \bm{q}/W$ is the recoil-corrected momentum transfer.
The overlap $\Phi$ between initial and final nuclear states is  
often called the bound-state wave function and includes the spectroscopic
factor.
According to standard distorted-wave theory \cite{Satchler83,Rawitscher97},
the wave function $\Psi^{(-)}$ appears with incoming boundary conditions
corresponding to incoming spherical waves in open channels of the  
$N+B$ system with an outgoing plane wave normalized to unit flux and is
related to standard scattering wave functions,$\Psi^{(+)}$, by time reversal.
Both wave functions are assumed to satisfy local single-nucleon Dirac 
equations of the form
\begin{subequations}
\label{eq:Dirac}
\begin{eqnarray}
\left[ \bm{\alpha} \cdot \bm{p} + \beta(m+S_b) + (V_b-E_b)\right] \Phi &=& 0 \\
\left[ \bm{\alpha} \cdot \bm{p} + \beta(m+S_c) + (V_c-E_c)\right] \Psi^{(+)} &=& 0
\end{eqnarray}
\end{subequations}
where $\Phi$ and $\Psi$ are four-component Dirac spinors, $S$ and $V$ are 
scalar and vector potentials, and the subscripts $b$ and $c$ distinguish 
between bound (initial) and continuum (final) states.

The bound-state spinor takes the form
\begin{equation}
\Phi_{\kappa m}({\bf r}) = \left(
\begin{array}{r}
f_{\kappa}(r) {\cal Y}_{\kappa m}(\hat{r}) \\
i g_{-\kappa}(r) {\cal Y}_{-\kappa m}(\hat{r})
\end{array} \right)
\end{equation}
where
\begin{equation} 
{\cal Y}_{\kappa m}(\hat{r}) = \sum_{s,m_s}
\langle \begin{array}{ll} \ell & \frac{1}{2} \\ \nu & m_s \end{array} |
\begin{array}{l} j \\ m \end{array} \rangle 
Y_{\ell\nu}(\hat{r})\chi_{m_s}
\end{equation}
is the spin spherical harmonic and where the orbital and total angular 
momenta are given by
\begin{subequations}
\begin{eqnarray}
\ell &=& S_\kappa(\kappa + \frac{1}{2}) - \frac{1}{2} \\
j &=& S_\kappa \kappa - \frac{1}{2}
\end{eqnarray}
\end{subequations}
with $S_\kappa=sign{(\kappa)}$.
The momentum distribution is then 
\begin{equation}
\rho(p) = \frac{1}{2\pi^2} \left( 
|\tilde{f}_{\kappa}(p)|^2 + |\tilde{g}_{\kappa}(p)|^2 \right)
\end{equation}
where
\begin{subequations}
\begin{eqnarray}
\tilde{f}_{\kappa}(p) &=& \int dr \; r^2  j_{\ell}(p r) f_{\kappa}(r) \\
\tilde{g}_{-\kappa}(p) &=& \int dr \; r^2  j_{\ell^\prime}(p r) g_{-\kappa}(r)
\end{eqnarray}
\end{subequations}
Similarly, the ejectile distorted wave is represented by
\begin{equation}
\Psi({\bf p},{\bf r}) = \sqrt{\frac{E_c + m}{2E_c}} 
\left( \begin{array}{r} 
\psi({\bf r}) \\  \zeta({\bf r}) \end{array} \right)
\end{equation}
where $\psi$ and $\zeta$ are two-component Pauli spinors and the
boundary conditions are suppressed.

It is often convenient to transform the two coupled first-order radial
Dirac equations into a single second-order equation 
\begin{equation}
 \left[ \nabla^2 + k^2 - 2\mu \left( U^C + U^{LS} \bm{L} \cdot 
\bm{\sigma} \right) \right] \xi = 0 
\end{equation}
where $\xi$ is a two-component Pauli spinor \cite{Sherif86,TIMORA,Kelly96}.
Here $k$ is the relativistic wave number,
$\mu$ is the nucleon mass for the bound state or the
relativistic reduced energy for the scattering state, 
and $U^C$ and $U^{LS}$ are central and spin-orbit potentials.
The corresponding Dirac wave functions are then given by
\begin{subequations}
\begin{eqnarray}
\Phi &=& \Omega_b \xi_b \\
\Psi &=& \Omega_c \xi_c
\end{eqnarray}
\end{subequations}
where
\begin{equation}
\label{eq:Omega}
\Omega(\bm{p},r) = 
\left( \begin{array}{c}
1 \\ \frac{\bm{\sigma \cdot p}}{(E+m) D(r)} 
\end{array} \right) D^{1/2}(r)
\end{equation}
is a spinor-distortion operator based upon relativistic effective mass
\begin{equation}
D(r) = 1 + \frac{S(r)-V(r)}{E+M}
\label{eq:D}
\end{equation}
in the presence of Dirac scalar and vector potentials, $S$ and $V$.
Therefore, the direct Pauli reduction method \cite{Hedayati-Poor95,Kelly99b}
is based upon a $2 \times 2$ current operator of the form
\begin{equation}
J^{\mu}(\bm{p}^\prime,\bm{p}) = 
\tilde{\Omega}_c(\bm{p}^\prime,r) \gamma^0 \Gamma^\mu \Omega_b(\bm{p},r)
\label{eq:J}
\end{equation}
that acts between the Pauli spinors for the relativized Schr\"odinger 
equations --- the lower components of the original Dirac spinors have
been incorporated within the operator.
Note that the transpose of the ejectile spinor distortion appears instead
of its Hermitian conjugate because of its incoming boundary conditions. 
Thus, one can identify two primary relativistic effects.
First, the Darwin potential $D(r)$ reduces the interior wave function 
and generally has a narrowing effect upon the momentum distribution.
Although the net effect is to increase spectroscopic factors fit
to missing momentum distributions, such effects are similar to those
of Perey nonlocality factors often used in nonrelativistic models
\cite{Udias95}.
Second, the presence of Dirac potentials produces dynamical enhancement
of the lower components.
Some observables, such as $A_{LT}$, are sensitive to this uniquely
relativistic effect through coupling between upper and lower components
mediated by the electromagnetic current operator.

Several approximations provide useful insight into the roles of various
relativistic effects.
The effective momentum approximation (EMA) replaces the momentum operators 
that appear in spinor-distortion operators by asymptotic kinematics.
Thus, the the initial momentum of the struck nucleon, $\bm{p}_b$, is
identified with the asymptotic missing momentum 
$\bm{p}_m=\bm{p}^\prime - \bm{q}$.
The so-called noSV approximation then uses
\begin{equation}
\label{eq:noSV}
\Omega(\bm{p},r) \longrightarrow 
\left( \begin{array}{c}
1 \\ \frac{\bm{\sigma \cdot p}}{E+m} 
\end{array} \right) D^{1/2}(r)
\end{equation}
to eliminate the dynamical enhancement of lower components.
Note, however, that the common Darwin factor is retained 
because it has an important effect upon the momentum distribution.
The combined EMA-noSV approximation \cite{Kelly96,Kelly99b} has the 
advantage that it can be used with two-component nonrelativistic wavefunctions 
without using $p/m$ expansions of the current operator.
Such wave functions are usually represented by 
\begin{equation}
\Phi(r) = P_b(r)^{1/2} \xi_b(r)
\end{equation}
where
\begin{equation}
P_b(r) = \left( 1 - \frac{\mu_b \beta_\text{NL}^2}{2} U^C_b(r) \right)^{-1}
\end{equation}
is a Perey nonlocality factor \cite{Perey62} based upon the central 
binding potential and where $\beta_\text{NL} \sim 0.85$ fm.
Thus, if one replaces the Darwin potential in the spinor distortion 
factor by the Perey factor, 
the EMA-noSV approach becomes practically identical to a conventional
NRDWIA calculation.
Furthermore, only the ejectile spin-orbit potential violates factorization of
RDWIA calculations based upon the EMA-noSV approximation.
A more detailed discussion of factorization in these approximations has been
given by Refs. \cite{Caballero98a,Vignote04}.

The EMA-SV approximation introduced in Ref. \cite{Kelly99b} also replaces 
momentum operators by asymptotic kinematics but retains the scalar and vector 
potentials in the spinor-distortion operator, thereby providing an estimate of 
the effects of dynamical enhancement of the lower components of the Dirac 
spinors by the mean field that is faster and numerically simpler than a full 
RDWIA calculation.
When using potentials defined for the Schr\"odinger equation, the Darwin
potential can be obtained from the spin-orbit potential according to
\cite{Jin94a}
\begin{equation}
\label{eq:DLS}
D(r) = \exp{
\left( 2\mu \int_r^\infty dr^\prime r^\prime \; U^{LS}(r^\prime) \right) }
\end{equation}
These representations of the Darwin potential are equivalent for RDWIA 
calculations for which $U^C$ and $U^{LS}$ are defined in terms of $S$ and
$V$ but are not necessarily equivalent for NRDWIA calculations where the
central and spin-orbit potentials are optimized independently.
Nevertheless, it is sometimes instructive to perform EMA-SV calculations
using Woods-Saxon wave functions fitted to $(e,e^\prime p)$ data using
NRDWIA methods.
However, the difference between $P(r)$ based upon $U^C$ and $D(r)$ based
upon $U^{LS}$ can affect the momentum distribution.

\begin{figure}
\centering
\includegraphics[angle=90,scale=0.35]{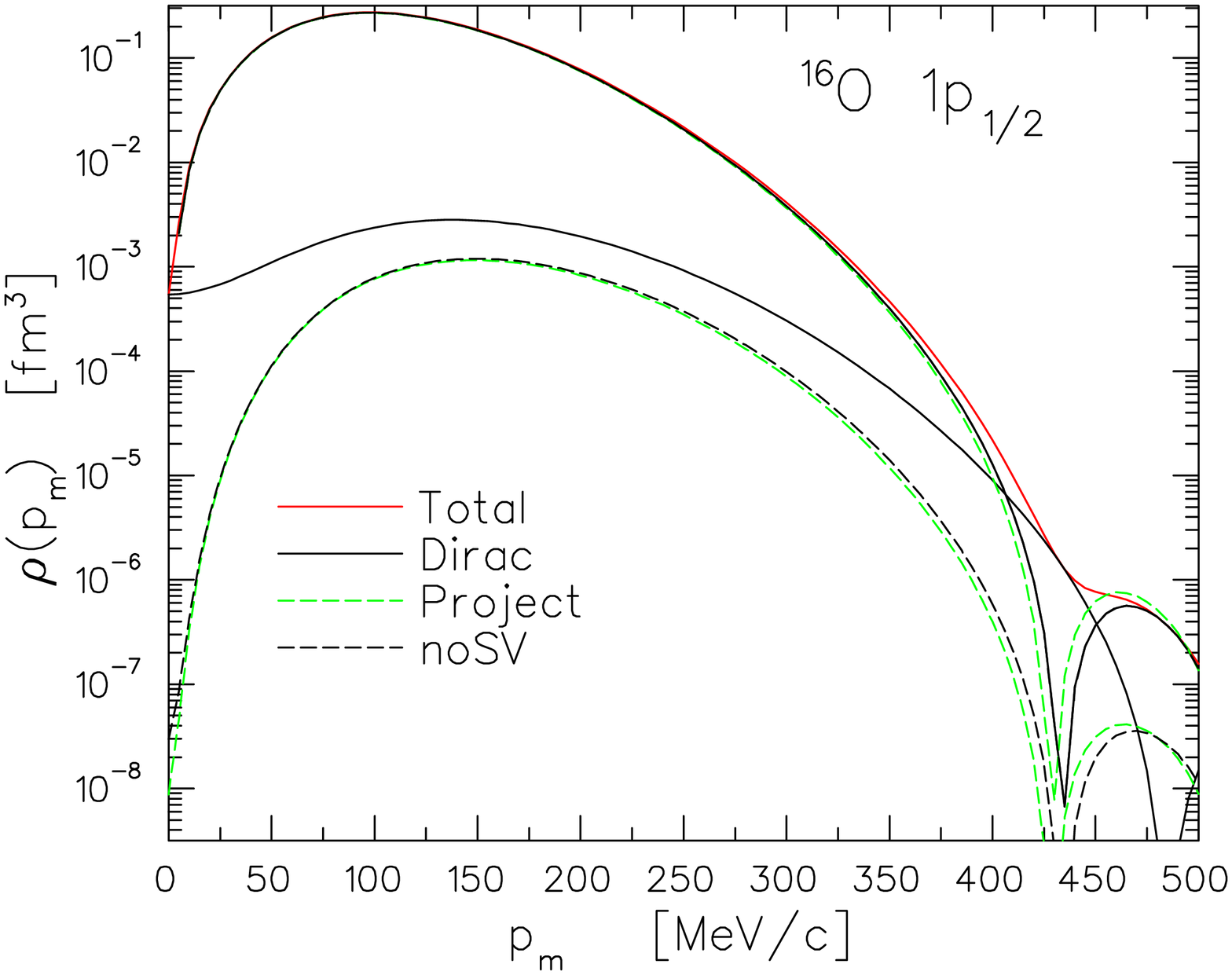}
\includegraphics[angle=90,scale=0.35]{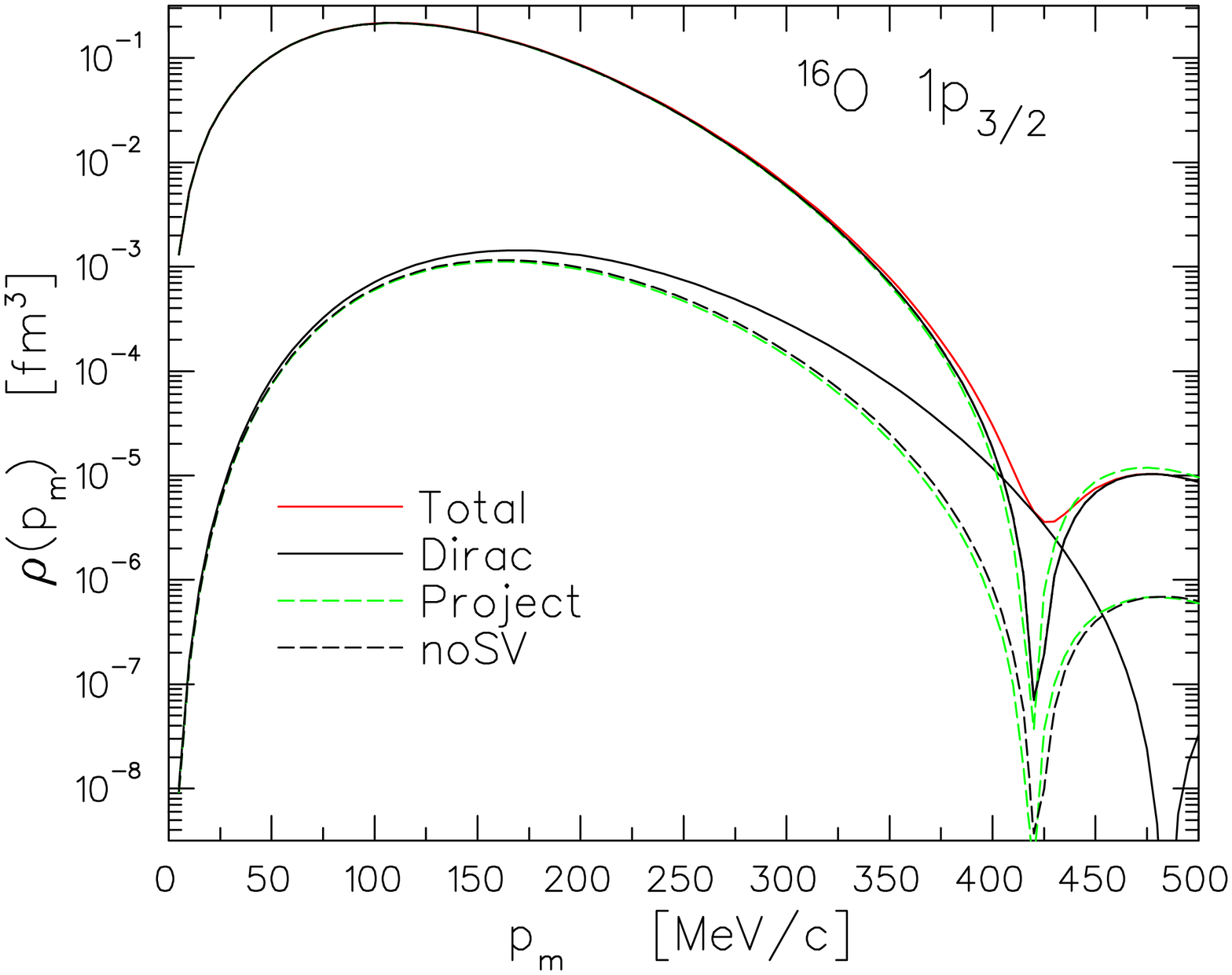}
\caption{(Color online)
Missing momentum distribution for $^{16}$O$(e,e^\prime p)(1p)^{-1}$.
The solid red curves show the total momentum distribution and solid black the 
contributions of upper and lower components of the Dirac wave functions.
The green dashed curves show the effect of projection on positive-energy states
while black dashed curves show the noSV approximation.
NLSH wave functions were used for this exercise.}
\label{fig:pm}
\end{figure}

The role of negative-energy contributions of the Dirac sea can be evaluated 
by applying the projection operator \cite{Udias01,Martinez04}
\begin{equation}
\Lambda_+(\bm{p}) = \frac{m+\not{p}}{2m}
\label{eq:Lambda}
\end{equation}
to $\Phi$, $\Psi$, or both.
Figure \ref{fig:pm} illustrates the effect of positive-energy projection
upon the momentum distributions for $1p$-shell knockout from $^{16}$O. 
The contribution of lower components is generally small, 
except near the diffraction minima of the upper component,
but can still be very important for observables, such as $A_{LT}$ that
emphasize interference between lower and upper components.
The negative-energy contribution is practically negligible for the upper 
component, but can be appreciable for the lower component.
The negative-energy contribution to the lower component of the $1p_{3/2}$ state
is small for low $p_m$ but increases with $p_m$, becoming dominant near the
diffraction minimum.
By contrast, the negative-energy contribution to the lower component of the 
$1p_{1/2}$ state is relatively strong for all $p_m$ --- 
in fact, it dominates for small $p_m$ because it has 
$\ell^\prime = \ell-1 = 0$ for $j=\ell - 1/2$.
For both states we find that the noSV approximation is very similar, but not
quite identical, to the result of positive-energy projection.
Therefore, we expect the noSV approximation to provide an accurate but
much faster estimate of the effect of positive-energy projection, at least for
the bound state.

\section{Results}

\subsection{$^{16}$O$(e,e^\prime p)$}
\label{sec:16O}

The left-right asymmetry in differential cross section is defined by
\begin{equation}
\label{eq:Alt}
A_{LT} = 
\frac{\sigma(\phi=0) - \sigma(\phi=\pi)}
     {\sigma(\phi=0) + \sigma(\phi=\pi)}
\end{equation}
where the subscript recognizes that this quantity is closely related 
to the longitudinal-transverse response function and
where the azimuthal angle $\phi=0$ corresponds to an ejectile momentum 
in the electron scattering plane between the beam direction and the 
momentum transfer. 
This quantity is especially sensitive to spinor distortion and 
has the advantages that it is independent of the spectroscopic factor
and for modest missing momentum is relatively insensitive to both the 
momentum distribution for the bound state and the optical potential 
for the ejectile.

\begin{figure}
\centering
\includegraphics[width=3.0in]{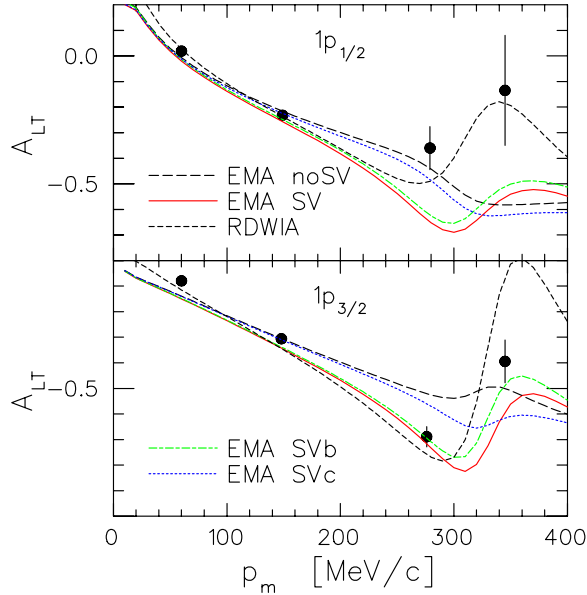}
\caption{(Color online) 
Left-right asymmetry for $^{16}$O$(e,e^\prime p)$ at $Q^2 = 0.8$ (GeV/$c$)$^2$.
Black dashed and red solid curves compare EMA-noSV and EMA-SV calculations, 
while green dash-dotted and blue dotted illustrate the effects of spinor
distortion on bound and ejectile wave functions, respectively.
Finally, the black short-dashed curves show full RDWIA calculations.}
\label{fig:EMAbc}
\end{figure}

Figure \ref{fig:EMAbc} examines the effect of spinor distortion for the
bound and/or ejectile wave functions in the context of EMA.
Except for small spin-orbit effects in the final-state, the EMA-noSV 
approximation nearly factorizes so that $A_{LT}$ is similar to that
for a free but moving nucleon.
Spinor distortion for the ejectile has relatively little effect at 
large $Q^2$ but is quite important for the bound state wave function,
strengthening $A_{LT}$ for $p_m \lesssim 300$ MeV/c and producing
pronounced oscillations at larger $p_m$ that can be attributed
to the breakdown of factorization.
Comparison between the black short-dashed and solid red curves shows that
EMA calculations are similar to those of the full RDWIA for 
$p_m \lesssim 250$ MeV/$c$, but tend to underestimate the oscillations
for larger $p_m$.

\begin{figure}
\centering
\includegraphics[width=3.0in]{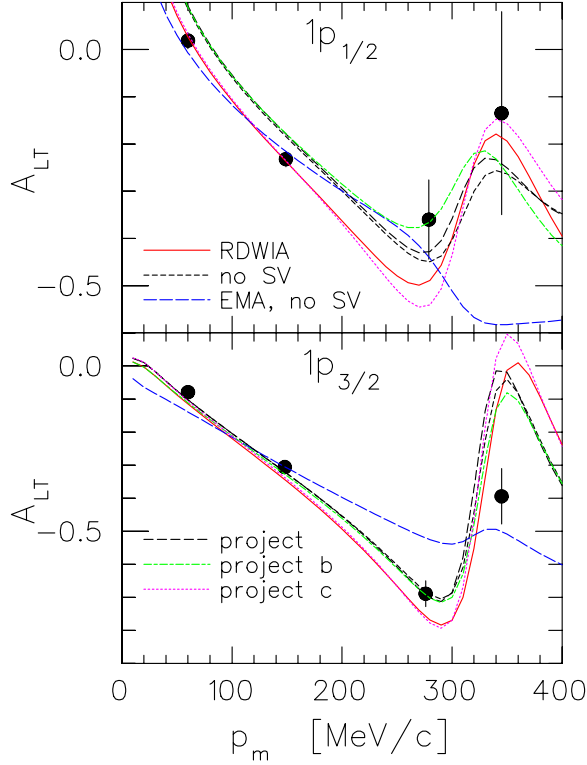}
\caption{(Color online) 
Left-right asymmetry for $^{16}$O$(e,e^\prime p)$ at $Q^2 = 0.8$ (GeV/$c$)$^2$.
The red solid curves show RDWIA, black short-dashed curves show 
noSV, blue long-dashed curves show EMA-noSV calculations.
Calculations with positive-energy projection are shown as green dash-dotted
curves for the bound nucleon, magenta dotted curves for the ejectile, or
black dashed curves for both.}
\label{fig:16O_project}
\end{figure}

Figure \ref{fig:16O_project} examines the sensitivity of $A_{LT}$ to the 
contribution of negative-energy states.
The differences between the EMA-noSV curves and the noSV curves show
that errors due to the effective momentum approximation can be appreciable 
even without spinor distortion.
The similarity between noSV curves (without EMA) and those that project upon 
positive-energy states shows that the dominant effect of spinor distortion
arises from coupling to the Dirac sea.
The fact that the noSV approximation is similar to positive-energy projection
was also observed in Fig. \ref{fig:pm} for the momentum distribution.
Positive-energy projection is also shown separately for the bound state (green
dash-dotted curves) and for the ejectile (magneta dotted curves).
Thus, comparing the curves labelled project $b$ or project $c$ with those
labelled RDWIA, 
we observe that the sensitivity of $A_{LT}$ to spinor distortion is 
greater for the bound state than for the ejectile and that positive energy 
projection is similar to the noSV approximation for both.
The difference between projected and RDWIA calculations shows that the 
sensitivity to the Dirac sea is greater for the $1p_{1/2}$ state, 
especially for small $p_m$.
That effect is subtle, but the low $p_m$ data do support the sea contribution.
Clearly it would be desirable to obtain more complete data for 
$p_m \lesssim 300$ MeV/$c$ where the single-nucleon knockout mechanism is 
most reliable.

\begin{figure}
\centering
\includegraphics[width=4in]{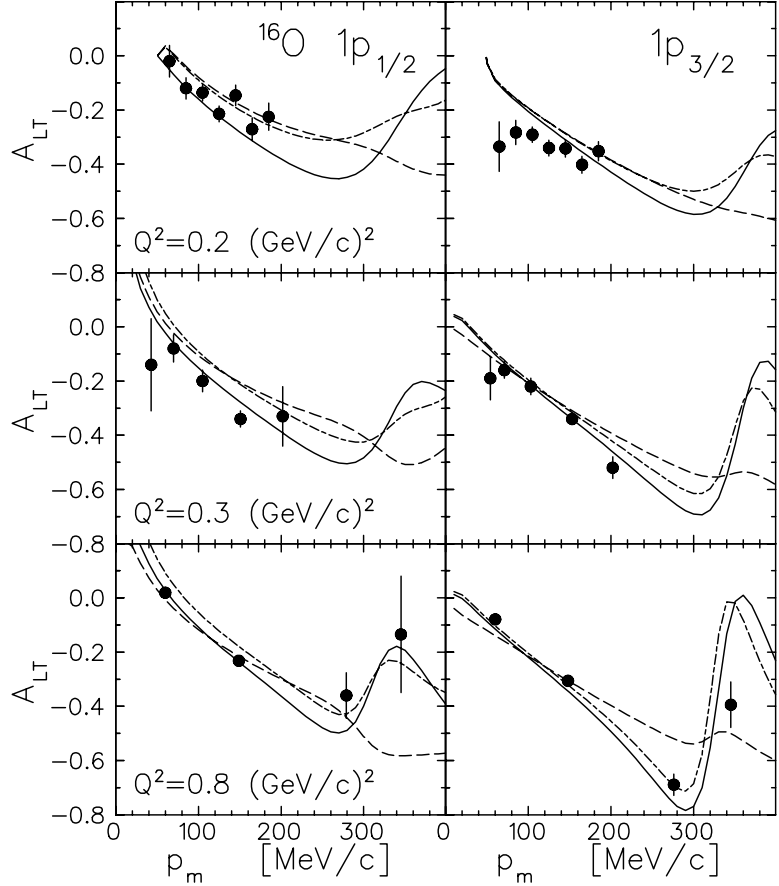}
\caption{
Dashed curves show EMA-noSV, solid curves RDWIA, and dash-dotted curves 
use positive-energy projection for $^{16}$O$(e,e^\prime p)$.}
\label{fig:16O_ALT_Q2}
\end{figure}

Figure \ref{fig:16O_ALT_Q2} surveys the $Q^2$ dependence of $A_{LT}$
for $^{16}$O$(e,e^\prime p)$.
The data for $Q^2 = 0.2$ are from Spaltro \etal \cite{Spaltro93} while
those for $Q^2 = 0.3$ (GeV/$c$)$^2$ are from 
Chinitz \etal \cite{Chinitz91}.
These calculations use NLSH wave functions, EDAD1 optical potential,
CC2 current operator, and Coulomb gauge.
The experimental asymmetry for the $1p_{1/2}$ state is slightly 
stronger than obtained from positive-energy states alone and is in
good agreement with the full RDWIA calculations.
The sensitivity to negative-energy states is smaller for the $1p_{3/2}$
state, but the RDWIA calculations are still in good agreement with 
the data for $Q^2 = 0.3$ and 0.8 (GeV/$c$)$^2$.
Although the calculations do not describe the $1p_{3/2}$ $A_{LT}$ data for 
$Q^2 = 0.3$ (GeV/$c$)$^2$ well, we are somewhat skeptical 
of the reliability of that particular data set because
Fissum \etal \cite{Fissum04} showed that there is an enhancement of 
this cross section for $50 < p_m < 120$ MeV/$c$ that cannot be reproduced 
by any of the many variations of the RDWIA model that were considered.
Therefore, we judge the overall agreement between RDWIA calculations 
and the $A_{LT}$ data for $^{16}$O$(e,e^\prime p)$ to support the 
participation of the Dirac sea in the bound-state wave function.
However, the data are not as precise nor the $p_m$ coverage as complete 
as we would like. 

\begin{figure}
\centering
\includegraphics[width=4in]{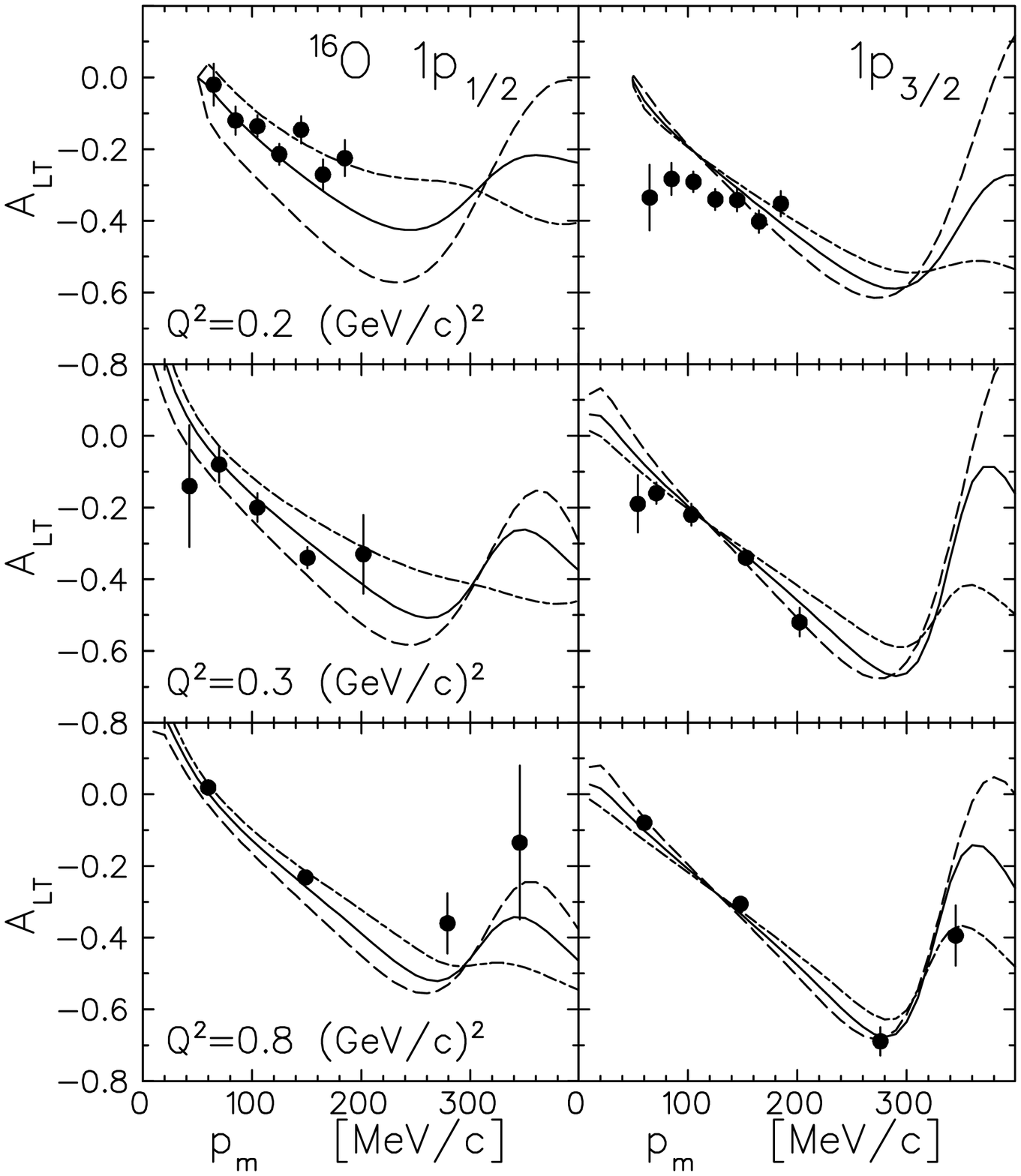}
\caption{
Dashed, solid, and dash-dotted curves show RDWIA calculations 
for $^{16}$O$(e,e^\prime p)$ using the CC1, CC2, and CC3 current
operators, respectively.}
\label{fig:16O_Gordon}
\end{figure}

Finally, Fig. \ref{fig:16O_Gordon} shows the sensitivity of $A_{LT}$ to
the Gordon ambiguity in the single-nucleon current operator.
All three operators are equivalent when the relationship between lower and
upper components is the same as for free nucleons, 
as in the EMA-noSV approximation, but spinor distortion breaks that
equivalence.
Thus, CC1 shows the greatest and CC3 the least sensitivity to 
spinor distortion.
This sensitivity is especially large for the $1p_{1/2}$ state and
decreases as $Q^2$ increases.
It is interesting to observe that dynamical relativistic effects are
most important at low $Q^2$.
The coupling of the ejectile to the Dirac sea is driven by $(S-V)/(E+m)$, 
which decreases as $E \sim Q^2/2m$ increases.
However, there is as yet no fundamental theory that distinguishes
between Gordon-equivalent forms of the single-nucleon current.
The available data for $^{16}$O$(e,e^\prime p)$ favor the intermediate
CC2 current operator.

\subsection{$^{12}$C$(e,e^\prime p)$}
\label{sec:12C}

Dutta \etal \cite{Dutta03} measured the reduced cross section for 
$^{12}$C$(e,e^\prime p)$ in quasiperpendicular kinematics at 
$Q^2 = 0.6$, 1.2, and 1.8 \GeVsq.
Although two beam energies are available for the first and third sets,
we consider only the higher energy and more forward electron-scattering
angle because the more complete coverage of the ejectile angle under those
conditions provides the left-right asymmetry.
The left-right asymmetry for reduced cross section is defined
\begin{equation}
\label{eq:alt}
a_{LT} = 
\frac{\sigma_\text{red}(\phi=0) - \sigma_\text{red}(\phi=\pi)}
     {\sigma_\text{red}(\phi=0) + \sigma_\text{red}(\phi=\pi)}
\end{equation}
and is small in PWIA because the intrinsic left-right asymmetry of {\it e-p} 
elastic scattering is removed by reduction of the differential cross section.
Note that we use a lower-case $a$ for the asymmetry in reduced cross
section and upper-case $A$ for the corresponding asymmetry in
differential cross section.

The data were presented in three bins: 
the lower bin with $15 \leq E_m \leq 25$ MeV is dominated by the
$1p_{3/2}$ contribution, the upper bin with $30 \leq E_m \leq 50$ MeV 
emphasizes the $1s_{1/2}$ contribution, while the semi-inclusive bin
with $10 \leq E_m \leq 80$ MeV also includes a significant continuum.
For the present purposes it is sufficient to treat the lower bin as
pure $1p_{3/2}$, the upper bin as pure $1s_{1/2}$, and the inclusive
bin as an incoherent mixture based upon the independent particle
shell model (IPSM).
The calculations for each orbital are based upon Dirac-Hartree 
wave functions and neglect the spreading with respect to missing energy.
We apply the same parametrization to the inclusive bin because we do
not have a detailed calculation of the continuum.
Although neither of the two narrow bins is pure and the inclusive bin 
includes continuum contributions, $a_{LT}$ for single-nucleon knockout
is rather insensitive to small deviations with respect to IPSM and 
we are more interested here in sensitivities to various aspects of the 
reaction model than in optimization of the spectral function.
A subsequent paper \cite{Kelly05b} will analyze the spectroscopic factors 
and nuclear transparency for these data.

\begin{figure}
\centering
\includegraphics[angle=90,width=5in]{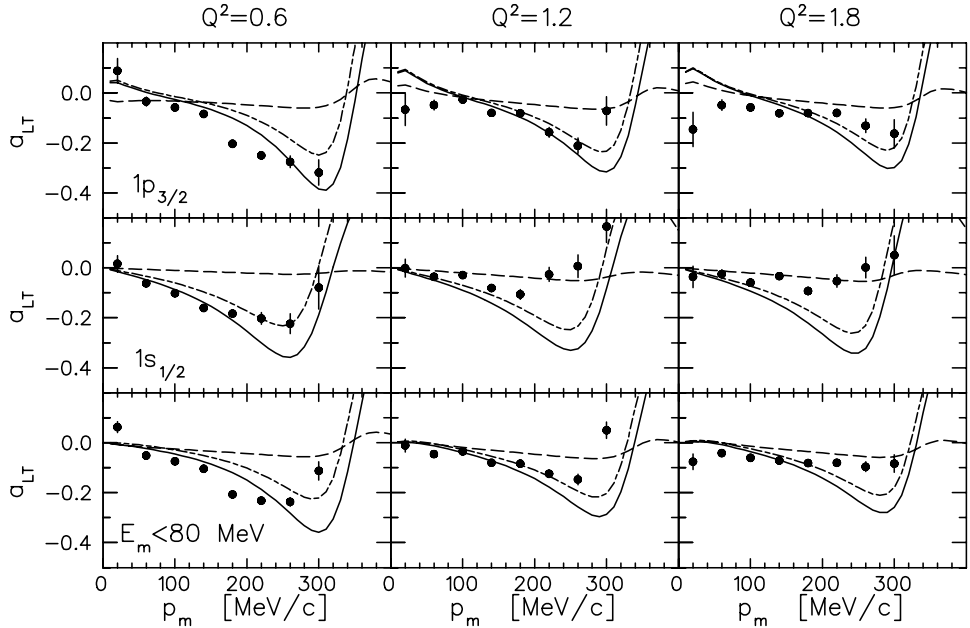}
\caption{
Dashed curves show EMA-noSV, solid curves RDWIA, and dash-dotted curves 
use positive-energy projection for $^{12}$C$(e,e^\prime p)$. 
The top row shows data for $15 \leq E_m \leq 25$ MeV, the middle row for
$30 \leq E_m \leq 50$ MeV, and the bottom row for $10 \leq E_m \leq 80$ MeV.}
\label{fig:12C_project}
\end{figure}

The sensitivity of $a_{LT}$ to the treatment of lower components is examined
in Fig. \ref{fig:12C_project}.
The dashed curves show that the left-right asymmetry for EMA-noSV is very
similar to that for RPWIA because the relationship between lower and upper
components is based upon asymptotic kinematics for free nucleons.
The very small variations of $a_{LT}$ are due to distortion effects that do 
not factorize completely.
The dash-dotted curves use wave functions projected onto positive-energy 
states that produce similar but somewhat smaller asymmetries than the full
calculation shown by solid curves.
The small differences between those curves is indicative of the contribution
of the Dirac sea, whose importance decreases slowly with increasing $Q^2$.

\begin{figure}
\centering
\includegraphics[angle=90,width=5in]{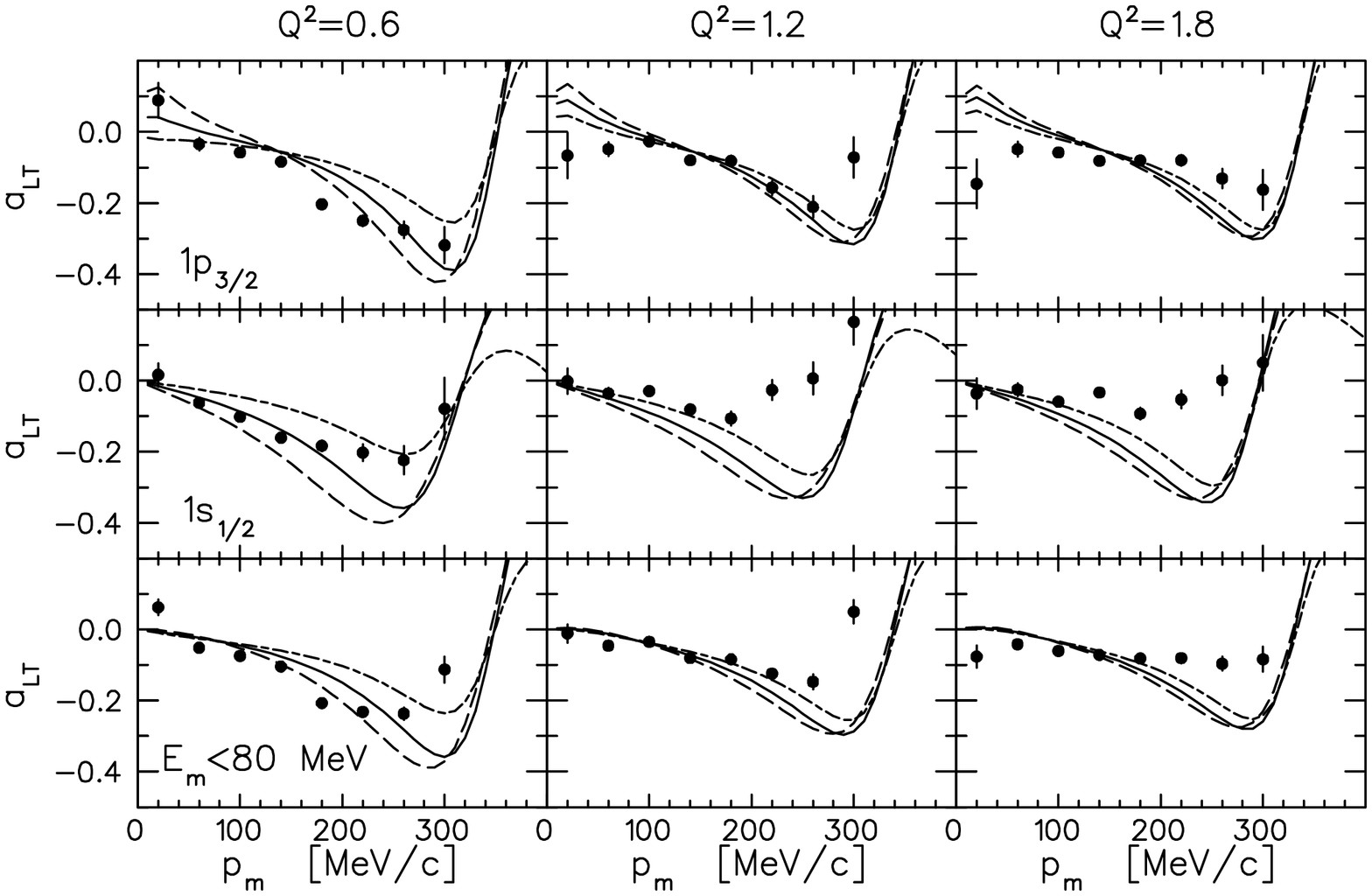}
\caption{
Dashed, solid, and dash-dotted curves show $^{12}$C$(e,e^\prime p)$ 
calculations using the CC1, CC2, and CC3 current operators.
All use NLSH wave functions, EDAD1 optical optical potentials, and Coulomb gauge.
The top row shows data for $15 \leq E_m \leq 25$ MeV, the middle row for
$30 \leq E_m \leq 50$ MeV, and the bottom row for $10 \leq E_m \leq 80$ MeV.}
\label{fig:12C_Gordon}
\end{figure}

The sensitivity of RDWIA calculations to the Gordon ambiguity in the 
single-nucleon current operator is displayed in Fig. \ref{fig:12C_Gordon}.
Clearly, the CC1 operator is most and the CC3 operator is least sensitive to
dynamical enhancement of the lower component of Dirac wave functions.
This sensitivity decreases fairly rapidly as $Q^2$ increases and has nearly
disappeared by 1.8 \GeVsq.
The data for lower $Q^2$ appear to favor the CC2 operator with intermediate
sensitivity to this effect.

\begin{figure}
\centering
\includegraphics[angle=90,width=5in]{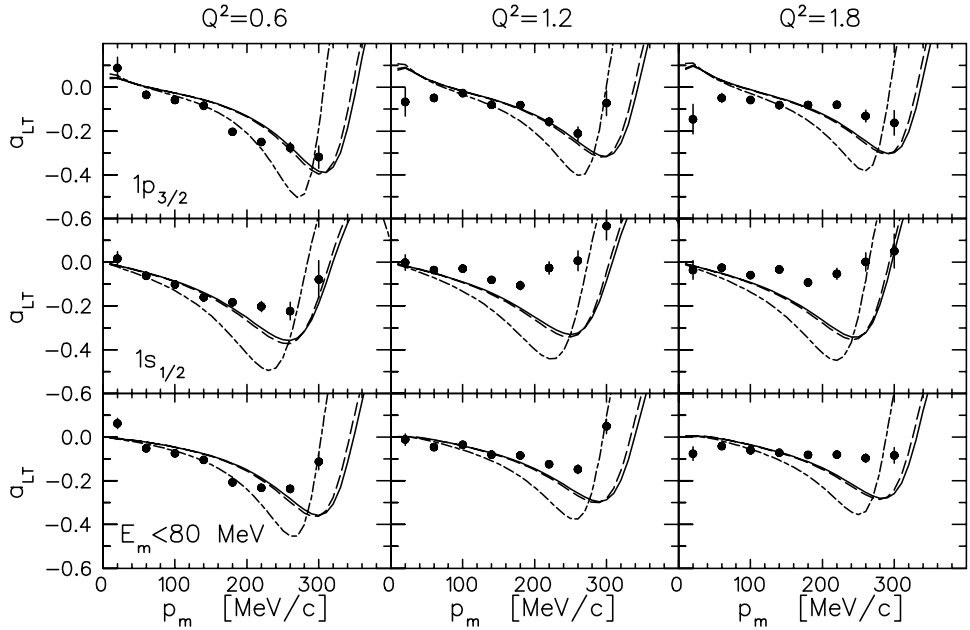}
\caption{
Solid, dashed and dash-dotted curves show $^{12}$C$(e,e^\prime p)$ 
calculations using NLSH, HS, and NIKHEF wave functions.
All use the CC2 current operator, Coulomb gauge, and EDAD1 optical
potentials.}
\label{fig:12C_wf_alt}
\end{figure}

The sensitivity to the overlap function is illustrated in 
Fig. \ref{fig:12C_wf_alt}.
The calculations shown for the $1p$ and $1s$ bins assume pure
$1p_{3/2}$ and $1s_{1/2}$ wave functions while those for the 
inclusive bin assume full occupancy for both shells.
The results for the two relativistic wave functions, NLSH and HS, are 
practically identical but the Woods-Saxon wave functions fitted by
van der Steenhoven \etal \cite{vdSteenhoven88a,vdSteenhoven88b} to 
data from NIKHEF give significantly different results:
the turnaround occurs earlier and $a_{LT}$ is stronger, especially
for the $1s$ state.
The data generally favor the relativistic bound-state wave functions,
although the $1s$ distributions tend to be flatter than those 
calculations as $Q^2$ increases.
To obtain relativistic calculations using Woods-Saxon wave functions,
the Perey factor was replaced by a Darwin factor computed from the
spin-orbit potential according to Eq. (\ref{eq:DLS}).
The same Darwin potential was also used for dynamical enhancement of
the lower component of the Dirac spinor according to Eq. (\ref{eq:Omega}).
Recognizing that EMA-noSV calculations are quite insensitive to the
upper components because the violation of factorization by ejectile
distortion is small, the main difference between $a_{LT}$ calculations
using relativistic or nonrelativistic wave functions can be attributed
to the difference between the $S-V$ potentials.
Although the $S-V$ potential for the bound state can be obtained from the 
spin-orbit potential in principle, in practice this quantity is poorly 
constrained by nonrelativistic $(e,e^\prime p)$ calculations and is 
usually chosen somewhat arbitrarily.
An advantage of the Dirac-Hartree approach is that the Darwin potential
emerges naturally during the optimization of the binding energy.

\begin{figure}
\centering
\includegraphics[angle=90,width=5in]{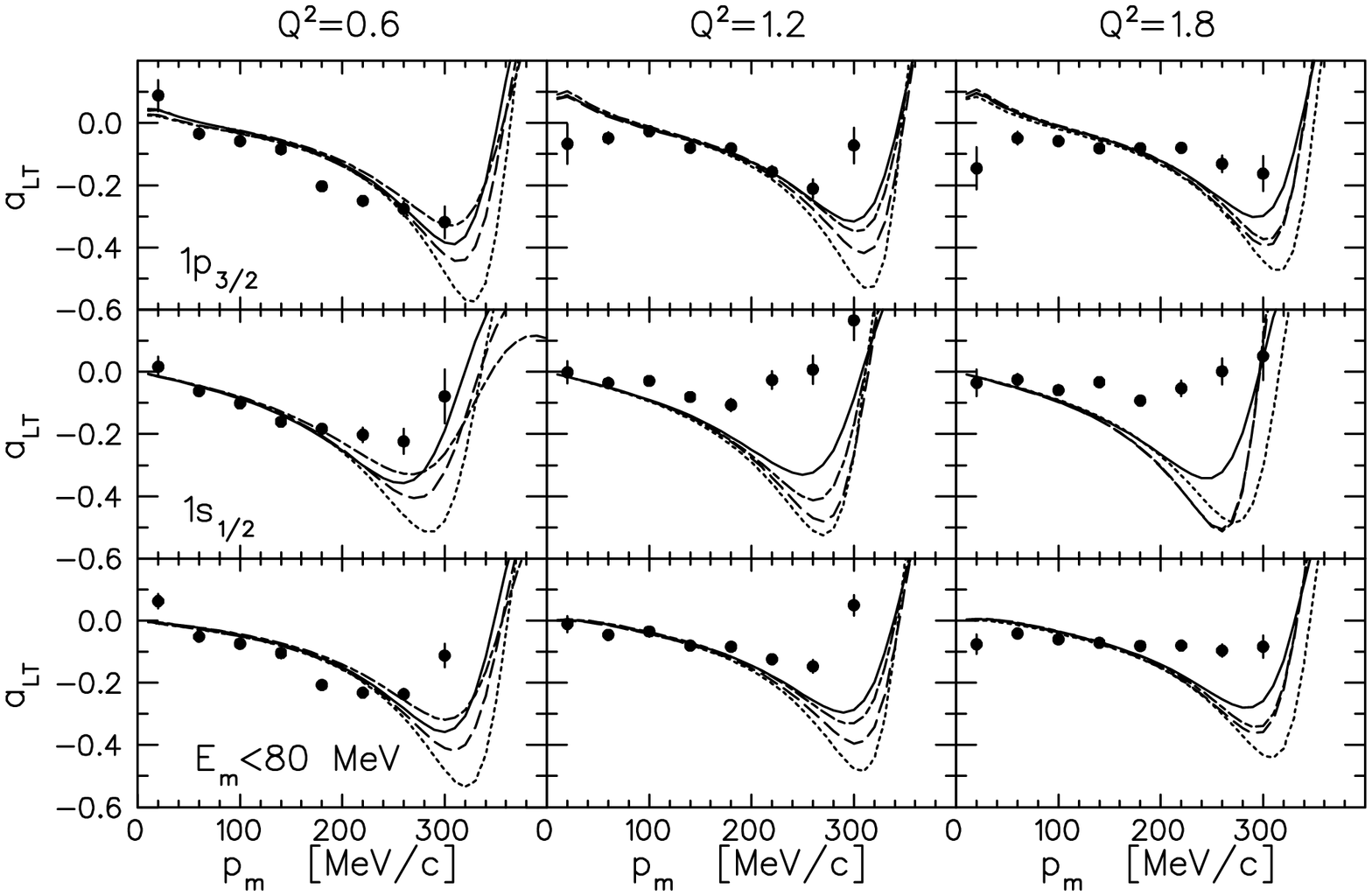}
\caption{
Dotted, solid, dashed and dash-dotted curves show $^{12}$C$(e,e^\prime p)$ 
calculations using the EDAIC, EDAD1, EDAD2, and EDAD3 optical potentials.
All use NLSH wave functions, CC2 current operator, and Coulomb gauge.}
\label{fig:12C_om}
\end{figure}

Figure \ref{fig:12C_om} shows that $a_{LT}$ for $p_m \lesssim 250$ MeV/$c$
is rather insensitive to the choice of optical potential.
Similar results were obtained using folding-model potentials based upon
the EEI or IA2 interactions \cite{Kelly94a}.
Therefore, $a_{LT}$ is more sensitive to the properties of the binding 
than the distorting potentials.
Slight shifts of the distorted momentum distribution produce some
sensitivity at larger $p_m$ to details of the optical potential, but 
there one might expect other effects, such as two-body currents or
channel coupling, to contribute also.

RDWIA calculations describe the $a_{LT}$ data for $Q^2 \lesssim 1.2$ 
(GeV/$c$)$^2$ relatively well, 
but the data for $Q^2 = 1.8$ (GeV/$c$)$^2$ show less variation with
$p_m$ than predicted.
The bin centered upon the $s$-shell also has a flatter $p_m$ distribution
for $Q^2 = 1.2$ (GeV/$c$)$^2$.
The flattening of $a_{LT}$ for $s$-shell and inclusive bins is probably 
due to continuum contributions that dilute the signal from single-nucleon
knockout --- although we do not have detailed calculations for the
continuum, there is little reason to expect multinucleon knockout to
retain the characteristic left-right asymmetry of single-nucleon knockout.
However, continuum contamination of the $p$-shell bin should be very
small because it lies below the two-nucleon emission threshold.
One possibility is that the ejectile spin-orbit potential may be 
too strong for $T_p \sim 900$ MeV because analyzing power data for 
proton elastic scattering are scarce near the upper limit of the energy
range used by Cooper \etal \cite{Cooper93}.
Alternatively, two-body currents may become important.

\clearpage

\section{Summary and Conclusions}
\label{sec:conclusions}

The most important difference between relativistic and nonrelativistic 
DWIA calculations of single-nucleon electromagnetic knockout is 
spinor distortion: the enhancement of lower components of Dirac spinors 
by the nuclear spin-orbit potential or, equivalently, the difference 
$S-V$ between scalar and vector potentials.
The left-right cross section asymmetry for quasiperpendicular
kinematics is especially sensitive to distortion of the bound-state
spinor.
We have used data for proton knockout from $^{16}$O and $^{12}$C to
evaluate the sensitivity of the left-right asymmetry to various aspects
of the RDWIA for $Q^2$ between about 0.2 and 1.8 (GeV/$c$)$^2$.
We find:

\begin{itemize}
\item
Negative-energy contributions are most important for $j<\ell$,
especially when $\ell=1$ permits negative-energy contributions for
$p_m=0$.
The $1p_{1/2}$ data for $^{16}$O$(e,e^\prime p)$ are described
significantly better by full RDWIA calculations than by either
EMA or positive-energy projection.
The noSV approximation is very similar to projection upon 
positive-energy states.

\item
The left-right asymmetry is more sensitive to the bound-state spin-orbit
potential than to variations of the optical potential.

\item
Among the most common Gordon-equivalent current operators, 
CC1 has the most and CC3 the least sensitivity to spinor distortion.
This sensitivity tends to decrease as $Q^2$ increases and is greater
for $1p_{1/2}$ than for $1p_{3/2}$ proton knockout.
The data for both $^{16}$O and $^{12}$C generally favor the intermediate 
choice, CC2. 

\item
The left-right asymmetry for $^{12}$C$(e,e^\prime p)$ with
$0.6 < Q^2 < 1.8$ (GeV/$c$)$^2$ is described fairly well by 
RDWIA calculations, but the data become flatter than
the calculations as $Q^2$ increases, especially for the $1s$ state.
This effect for the $1p$ state might indicate that the spin-orbit
potential in global optical potentials from Dirac phenomenology
is somewhat too strong for $T_p > 600$ MeV.
The more pronounced flattening for the $1s$ state might indicate
greater contamination by multinucleon continuum.

\end{itemize}

\begin{acknowledgments}
We thank Dr. Dutta for tables of the $a_{LT}$ data, Dr. Ud\'ias for tables
of the NLSH and HS wave functions, and Dr. Lapik\'as for the parameters
of the NIKHEF wave functions.
The support of the U.S. National Science Foundation under grant PHY-0140010
is gratefully acknowledged.
\end{acknowledgments}


\end{document}